\documentclass[
]{ceurart}

\usepackage{listings}
\begin{document}

\copyrightyear{2026}
\copyrightclause{Copyright for this paper by its authors.
  Use permitted under Creative Commons License Attribution 4.0
  International (CC BY 4.0).}

\conference{Proceedings of the Eighth International Workshop on Automated Semantic Analysis of Information in Legal Text (ASAIL 2026), 12 June 2026, Singapore}

\title{Prompting Beats Fine-Tuning: Generative Expected Value Scoring for Statutory Term Retrieval}


\author[1]{Alvin Wang}[email=alvinw2@andrew.cmu.edu]
\author[1]{Jaromir Savelka}[email=jsavelka@andrew.cmu.edu, orcid=0000-0002-3674-5456]
\address[1]{Carnegie Mellon University, 5000 Forbes Ave, Pittsburgh, PA 15213, USA}
\begin{abstract}
Legal concepts in statutes are often expressed using vague terms, and practitioners frequently turn to case law to interpret them. We study the task of ranking case-law sentences by their usefulness for explaining a concept or target statutory term, using an established dataset of 26,959 sentences covering 42 U.S. Code concepts labeled into four explanatory-value categories. We compare two families of methods: (i) supervised fine-tuning of encoder-only models (ModernBERT) and (ii) zero-shot prompting of decoder-only models. We show that across all concepts and standard NDCG cutoffs, ModernBERT largely matches earlier BERT-family baselines. In contrast, prompting decoder-only models achieves the strongest overall effectiveness, with our best system surpassing all previously reported state-of-the-art results on this task.
\end{abstract}


\maketitle

\section{Introduction}
Understanding laws may be challenging because they need to communicate general standards and refer to classes of persons, acts, things, and circumstances \cite[p. 124]{hart1994concept}. Therefore, legislators must use vague~\cite{endicott2000vagueness}, open textured \cite{hart1994concept} terms, abstract standards \cite{endicott2014law}, principles, and values \cite{daci2010legal}. Understanding of any provision of law may depend on understanding the meaning of a term mentioned within. Potential doubts about its meaning may be removed by explanation or interpretation \cite{maccormick1991interpreting}. When searching a database of legal documents, a lawyer may retrieve many short text snippets (sentences) that mention a particular term. Some of the sentences are most likely useful for explaining the term but others may have very little value. Manually reviewing all the snippets is labor intensive because they may often be retrieved in thousands or more.

In this work, we focus on the task of ranking text snippets (sentences) more highly that are useful for interpretation or explanation of a selected term as defined in \cite{savelka2021discovering}. These may include definitional sentences, sentences that state explicitly in a different way what the term means or state what it does not mean, sentences that provide an example, instance, or counterexample, and sentences that show how a court determines whether something is such an example.

This paper substantially extends our previously published demonstration paper \cite{wang2026explanatory}, which focused on the ranking task and offered the initial empirical comparison of encoder-only and decoder-only language models. That study produced two initial findings: (i) more modern fine-tuned encoder models perform comparably to earlier BERT-family baselines, and (ii) decoder-only prompting, particularly with GPT models, achieves competitive performance without any task-specific fine-tuning.

Building on those findings, the present work makes the following contributions: we (i) extend the model comparison beyond proprietary OpenAI systems to include open-weight models,
(ii) provide a more detailed experimental setup and evaluation protocol, (iii) analyze performance across multiple query regimes (small/large, sparse/dense), and (iv) offer a deeper investigation into the role of context and model architecture in explanatory sentence ranking. Results reported in this paper are state-of-the-art.

\section{Related Work}
In prior work, researchers employed a variety of traditional information retrieval measures and their combinations, e.g., BM25, novelty, topic modeling \cite{savelka2019improving,savelka2020discovering,savelka2021legal}. These turned out to be successful in finding documents or their parts that are likely to contain useful sentences. However, they did not perform a finer-grained evaluation of the sentences contained in those parts of the document. Using learning-to-rank approaches on hand-made features led to only moderate improvements \cite{savelka2016extracting,savelka2020learning}. They also showed that transformer-based pre-trained models (BERT family) were capable of such fine-grained evaluation by learning to detect sophisticated semantic features in sentences themselves and in their relationships to the explained terms \cite{savelka2021discovering}. % We need to add the Polish paper.
Recently, DeBERTa-based approaches (fine-tuning) and QWEN-2.5 (prompting) improved on previous work reporting current state-of-the-art results \cite{libal2025manual}. In this work, we largely confirm the results of the prior work and introduce novel methods that match the state-of-the-art in zero-shot settings.

\section{Dataset}
The dataset was originally introduced by \cite{savelka2021discovering}. It was constructed by selecting 42 legal concepts from provisions of the U.S. Code and querying the Caselaw Access Project corpus for judicial sentences containing those concepts. The resulting collection comprises 26,959 sentences spanning 20 areas of U.S. legal regulation. Each sentence was independently assessed by law student annotators with respect to its usefulness for explaining the meaning of the target legal concept. Following the annotation guidelines proposed by \cite{savelka2021discovering}, law students classified the sentences into four categories according to their usefulness for explaining the terms:

\begin{enumerate}
    \item \textbf{High value} -- sentences the goal of which is to elaborate on the meaning of the term.
    \item \textbf{Certain value} -- sentences that provide grounds to draw some conclusions about the meaning of the term.
    \item \textbf{Potential value} -- sentences providing additional information over what is known from the provision.
    \item \textbf{No value} -- sentences that do not provide any additional information.
\end{enumerate}
 These categories reflect progressively lower levels of explanatory value, ranging from explicit elaborations of a concept's meaning to sentences providing no additional interpretive information. The annotation quality was assessed using Krippendorff's alpha \cite{krippendorff2011computing}, with two legally trained annotators achieving an agreement score of 0.79, indicating substantial inter-annotator reliability.
% \noindent The inter-annotator agreement was ($\alpha=0.79$) \cite{krippendorff2011computing}.

\begin{figure}[]
    \centering
    \includegraphics[width=0.47\textwidth]{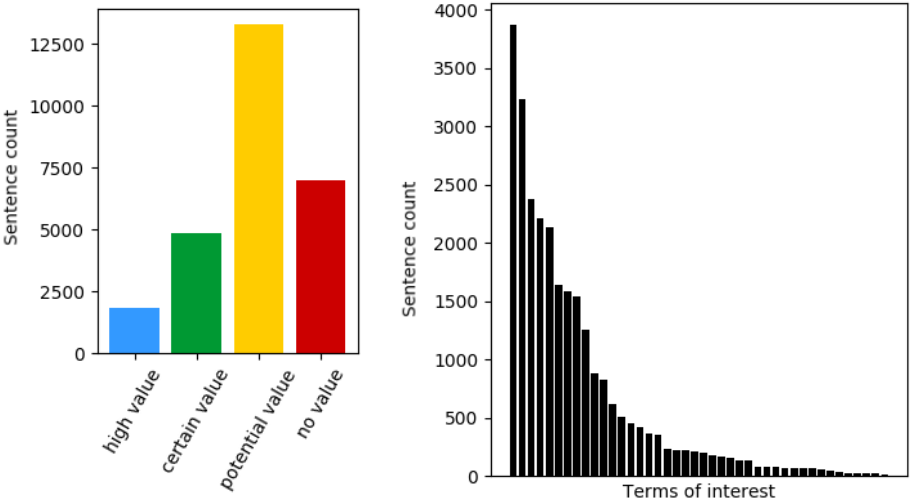}
    \caption{The graph on the left shows the distribution of the labels. The graph on the right presents the distribution of the number of sentences retrieved for each term. The graph is reproduced from \cite{savelka2021discovering} with the permission of the authors.}
    \label{fig:data_set_overview}
\end{figure}

Figure \ref{fig:data_set_overview} shows the overall distribution of the labels. The figure shows that the less relevant categories (`no value' and `potential value') are dominant. Some of the terms contain quite a considerable number of more relevant sentences while others are significantly more limited in this respect.

\section{Retrieval Methods}
We investigate two families of approaches: (i) \emph{fine-tuning} encoder-only language models; and (ii) \emph{zero/-shot} prompting of decoder-only language models.

% \subsubsection{Input Representation}
% All encoder-only variants follow the standard ModernBERT classification setup, in which a single serialized input sequence is constructed and a sentence-level relevance label is predicted. Variants differ only in the contextual information included alongside the candidate sentence.

\subsection{Fine-tuning of Encoder-only Language Models}
In this set of experiments, we employ ModernBERT~\cite{warner2025smarter} which exhibits state-of-the-art results on a large pool of evaluations encompassing both single- and multi-vector retrieval on different domains. We re-implement setups presented in prior work \cite{savelka2021discovering} that relied on earlier BERT \cite{devlin2019bert} and RoBERTa~\cite{liu2019roberta} models, and experiment with additional novel setups. Compared to earlier models, ModernBERT incorporates improvements for longer context handling and inference efficiency~\cite{warner2025smarter}. It has been trained on approximately 2 trillion tokens and natively supports sequence lengths up to 8{,}192 tokens. Architecturally, it replaces absolute positional embeddings with rotary positional embeddings (RoPE)~\cite{su2024roformer}, and uses an efficiency-oriented attention pattern, enabling long-context inference at substantially lower cost than full attention in every layer.

We implement the following six retrieval approaches (first three from prior work; additional three ours):

\begin{enumerate}
    \item \textbf{snt} -- Fine-tuning to predict the explanatory value of a retrieved sentence based on the sentence itself. %This setting evaluates the extent to which sentence-only cues alone are helpful in relevance estimation.
    \item \textbf{qry2snt} -- The model predicts the value of the sentence based on the sentence and the term of interest. % This configuration explicitly conditions relevance judgments on the query, aligning more closely with the dataset’s annotation protocol.
    \item \textbf{sp2snt} -- Prediction is based on the full statutory provision that contains the term and the retrieved sentence. % This allows the model to assess whether the sentence contributes information beyond the source text itself and to penalize near-duplicates or paraphrases..
    \item \textbf{sp+qry2snt} -- We extend \textbf{sp2snt} by marking occurrences of the term in the provision with special tokens. %Highlighting the concept guides the model’s attention to the relevant portion of the statute while preserving surrounding legal context.
    \item \textbf{sp2ctx} -- We extend \textbf{sp2snt} by embedding the retrieved sentence within a paragraph it comes from.% This setting captures explanatory signals that arise from local narrative structure rather than sentence-level content alone.
    \item \textbf{sp2snt+ctx} -- We combine standalone and context embedded views of the retrieved sentence. %The sentence is presented both independently and within its paragraph context, enabling the model to score sentence-level relevance while simultaneously leveraging discourse-level information.
\end{enumerate}

\begin{table}
\caption{Token representations and max token limits used for ModernBERT-based approaches. Here, $snt$ denotes the candidate sentence, $qry$ the legal term of interest, $sp$ the source statutory provision, and $ctx$ the context paragraph.}
\label{tab:token_representations}
\setlength{\tabcolsep}{4.5pt}
\centering
\small
\begin{tabular}{l l c}
\toprule
\textbf{Method} & \textbf{Token Representation} & \textbf{Max} \\
\midrule
snt        & $\texttt{[CLS]} \; snt \; \texttt{[SEP]}$                                                                       & 512 \\
qry2snt    & $\texttt{[CLS]} \; qry \; \texttt{[SEP]} \; snt \; \texttt{[SEP]}$                                              & 512 \\
sp2snt     & $\texttt{[CLS]} \; sp \; \texttt{[SEP]} \; snt \; \texttt{[SEP]}$                                               & 840 \\
sp+qry2snt & $\texttt{[CLS]} \; sp{\langle qry \rangle} \; \texttt{[SEP]} \; snt \; \texttt{[SEP]}$                          & 840 \\
sp2ctx     & $\texttt{[CLS]} \; sp \; \texttt{[SEP]} \; ctx{\langle snt \rangle} \; \texttt{[SEP]}$                          & 1280 \\
sp2snt+ctx & $\texttt{[CLS]} \; sp \; \texttt{[SEP]} \; snt \; \texttt{[SEP]} \; ctx{\langle snt \rangle} \; \texttt{[SEP]}$ & 1440 \\
\bottomrule
\end{tabular}

\end{table}

\noindent Table \ref{tab:token_representations} provides additional details.
% \noindent\textbf{Sentence (snt).}
% \noindent\textbf{Query--Sentence (qry2snt).}
%\noindent\textbf{Source Provision--Sentence (sp2snt).}
% \noindent\textbf{Source Provision + Query--Sentence (sp+qry2snt).}
% \noindent\textbf{Source Provision--Context (sp2ctx).}
% \noindent\textbf{Source Provision--Sentence + Context (sp2snt+ctx).}
% \vspace{-\baselineskip}

\subsection{Prompting of Decoder-only Models}
We prompt decoder-only models to predict sentence relevance, employing a compacted version of the annotation guidelines used to construct the original dataset. We experiment with a diverse set of models that span proprietary and open-weight systems across multiple providers.

From OpenAI, we include \textbf{GPT-4o} \cite{hurst2024gpt}, an autoregressive omni model that accepts any combination of text, audio, image, and video as input; \textbf{GPT-OSS-120B} \cite{openai2025gptoss}, OpenAI's open-weight model; \textbf{GPT-5.2} \cite{openai2025gpt52} and \textbf{GPT-5.4} \cite{openai2025gpt54}, representing OpenAI's most capable model series. From Meta, we include \textbf{LLaMA-3.3-70B} \cite{grattafiori2024llama3} and \textbf{LLaMA-4-Scout-17B} \cite{meta2025llama4}, both open-weight models. Finally, we include \textbf{Qwen-3-32B} \cite{qwen2025qwen3}, an open-weight model from Alibaba.

Based on findings from our prior work, we adopt the \textbf{Probabilities} prompting approach, which consistently outperformed zero-shot, few-shot, and large-context variants. Rather than generating a discrete relevance label, the model outputs a probability distribution over the four classes. The final ranking score is the expected value (EV) of the distribution, calculated as $\sum_{i=0}^{3} i \cdot P(\text{class}=i)$. Each user message includes the legal term, the source provision, and a batch of 25 retrieved sentences, with temperature set to 0.0 for near-deterministic output.
% \noindent\textbf{Pointwise Classification.}

% \noindent\textbf{Listwise Ranking.}
% For list-wise retrieval, the model is given the full document context associated with a legal concept, with candidate sentences explicitly marked in situ. Rather than scoring sentences independently, the model is instructed to produce a ranked list of sentence identifiers. This formulation enables direct comparison among candidates within their natural document context, capturing global relevance signals that are typically inaccessible to pointwise methods.

\section{Experiments}
% \subsection{Experimental Design}
% We evaluate two families of retrieval approaches, summarized in Table~\ref{tab:full_results}. 

% \begin{table}[h]
% \centering
% \small
% \begin{tabular}{l l l l}
% \toprule
% \textbf{Method} & \textbf{Context} & \textbf{Paradigm} & \textbf{Output} \\
% \midrule
% ModernBERT & Sentence-Centric & Supervised & Label (0--3) \\
% Zero-shot & Sentence, Statute & Pointwise & Label (0--3) \\
% Few-shot & Sentence, Statute & Pointwise & Label (0--3) \\
% Probabilistic & Sentence, Statute & Pointwise & EV (0--3) \\
% Large-context & Full Document & Listwise & Rank \\
% \bottomrule
% \end{tabular}
% \caption{Summary of ranking methods evaluated. EV denotes the expected value of the relevance label distribution.}
% \label{tab:methods}
% \end{table}

\subsection{Finetuning ModernBERT}
We adopt the experimental design from \cite{savelka2021discovering} by partitioning the data set into six disjoint folds using stratified sampling. In each run, the model is trained on sentences associated with four folds, validated on a fifth fold, and evaluated on a held-out test fold. Models are trained for 10 epochs with a batch size of 16 and a learning rate of $2\times10^{-5}$. Training is performed using the AdamW optimizer \cite{loshchilov2017decoupled} with a weight decay of 0.01, together with a linear learning-rate schedule without warm-up. Best model selection is based on the NDCG@100 computed on the validation fold, and only the best-performing checkpoint is retained for evaluation.

For the sentence-only settings ($snt$ and $qry2snt$), we use a 512-token limit to match prior encoder-only baselines and avoid attributing gains merely to longer context windows \cite{devlin2019bert,lin2020pretrained}. This limit is sufficient for nearly all candidate sentences: under the ModernBERT tokenizer, the mean sentence length is 63.6 tokens and the 95th percentile is 147 tokens. For settings that include the statutory provision, we increase the limit to 840 tokens, since the longest provision contains 321 tokens. Paragraph-context variants use larger limits of 1{,}280 and 1{,}440 tokens to accommodate the 250-token local context window and, in the latter case, the duplicated standalone sentence representation.

\subsection{Inference Infrastructure and Reliability}

To evaluate the performance of our systems, we submit queries across all 42 statutory terms using the \verb|openai| Python library, which provides a unified interface to both OpenAI's REST API and OpenAI-compatible endpoints (e.g., Groq). Queries are dispatched asynchronously via \verb|AsyncOpenAI|, with a semaphore limiting concurrency to prevent rate-limit violations. 

To ensure robustness against LLM formatting errors, we employ a retry mechanism (up to 5 attempts) with Regex-based preprocessing and strict JSON parsing via a custom \verb|extract_json_array| function. This handles common hallucinations such as spelled-out decimal digits (e.g., \texttt{"0. nine"} $\rightarrow$ \texttt{0.9}), invalid leading zeros, unclosed \verb|| tags from reasoning models, and dict-wrapped arrays.

\begin{table}[t]
\caption{NDCG@10 and NDCG@100 across small/large and sparse/dense query subsets, combining prior work with ModernBERT and GPT-based methods.}
\label{tab:full_results}
\centering
\small
\setlength{\tabcolsep}{4pt}
\resizebox{\textwidth}{!}{
\begin{tabular}{l cc cc cc cc cc}
\toprule
 & \multicolumn{2}{c}{SmSp} 
 & \multicolumn{2}{c}{SmDs} 
 & \multicolumn{2}{c}{LgSp} 
 & \multicolumn{2}{c}{LgDs} 
 & \multicolumn{2}{c}{Overall} \\
Method
 & @10 & @100 
 & @10 & @100 
 & @10 & @100 
 & @10 & @100 
 & @10 & @100 \\
\midrule
BERT snt \cite{savelka2021discovering}
 & .45 $\pm$ .13 & .66 $\pm$ .24
 & .67 $\pm$ .02 & .83 $\pm$ .04
 & .39 $\pm$ .32 & .37 $\pm$ .36
 & .82 $\pm$ .16 & .75 $\pm$ .06
 & .59 $\pm$ .19 & .70 $\pm$ .22 \\

BERT qry2snt \cite{savelka2021discovering}
 & .41 $\pm$ .19 & .66 $\pm$ .23
 & .73 $\pm$ .10 & .86 $\pm$ .03
 & .48 $\pm$ .68 & .43 $\pm$ .53
 & .98 $\pm$ .03 & .81 $\pm$ .13
 & .64 $\pm$ .30 & .73 $\pm$ .24 \\

BERT sp2snt \cite{savelka2021discovering}
 & .53 $\pm$ .27 & .78 $\pm$ .17
 & .67 $\pm$ .13 & .82 $\pm$ .06
 & .71 $\pm$ .23 & .48 $\pm$ .26
 & .95 $\pm$ .07 & .84 $\pm$ .02
 & .68 $\pm$ .21 & .77 $\pm$ .17 \\

DeBERTa-large (v3) \cite{libal2025manual}
 & -- & --
 & -- & -- 
 & -- & -- 
 & -- & -- 
 & \textbf{.79} & .79\\

Qwen2.5Instruct 72B \cite{libal2025manual}
 & -- & -- 
 & -- & -- 
 & -- & -- 
 & -- & -- 
 & .78 & \textbf{.85}\\

\midrule
ModernBERT snt
 & .47 $\pm$ .18 & .73 $\pm$ .16
 & .61 $\pm$ .09 & .80 $\pm$ .07
 & .40 $\pm$ .26 & .48 $\pm$ .18
 & .86 $\pm$ .13 & .73 $\pm$ .10
 & .58 $\pm$ .21 & .72 $\pm$ .16 \\

ModernBERT qry2snt
 & .54 $\pm$ .15 & .75 $\pm$ .12
 & .64 $\pm$ .14 & .81 $\pm$ .08
 & .55 $\pm$ .15 & .50 $\pm$ .15
 & .78 $\pm$ .28 & .72 $\pm$ .16
 & .62 $\pm$ .18 & .74 $\pm$ .15 \\

ModernBERT sp2snt
 & .62 $\pm$ .17 & .81 $\pm$ .11
 & .67 $\pm$ .13 & .82 $\pm$ .09
 & .53 $\pm$ .17 & .51 $\pm$ .18
 & .82 $\pm$ .17 & .78 $\pm$ .11
 & .66 $\pm$ .17 & .77 $\pm$ .15 \\

ModernBERT sp+qry2snt
 & .47 $\pm$ .12 & .73 $\pm$ .11
 & .55 $\pm$ .11 & .78 $\pm$ .11
 & .34 $\pm$ .25 & .32 $\pm$ .22
 & .65 $\pm$ .08 & .59 $\pm$ .04
 & .51 $\pm$ .16 & .67 $\pm$ .20 \\

ModernBERT sp2ctx
 & .49 $\pm$ .10 & .76 $\pm$ .11
 & .63 $\pm$ .16 & .81 $\pm$ .10
 & .51 $\pm$ .27 & .49 $\pm$ .18
 & .74 $\pm$ .18 & .72 $\pm$ .11
 & .59 $\pm$ .18 & .74 $\pm$ .16 \\

ModernBERT sp2snt+ctx
 & .46 $\pm$ .12 & .72 $\pm$ .15
 & .64 $\pm$ .16 & .81 $\pm$ .12
 & .28 $\pm$ .25 & .35 $\pm$ .21
 & .78 $\pm$ .05 & .67 $\pm$ .09
 & .56 $\pm$ .21 & .70 $\pm$ .20 \\

\midrule

LLaMA-3.3-70B
 & .62 $\pm$ .14 & .81 $\pm$ .10
 & .69 $\pm$ .17 & .86 $\pm$ .08
 & .28 $\pm$ .19 & .45 $\pm$ .16
 & .46 $\pm$ .32 & .63 $\pm$ .15
 & .58 $\pm$ .24 & .75 $\pm$ .18 \\

LLaMA-4-Scout-17B
 & .51 $\pm$ .13 & .78 $\pm$ .09
 & .68 $\pm$ .13 & .85 $\pm$ .07
 & .44 $\pm$ .14 & .49 $\pm$ .15
 & .65 $\pm$ .20 & .65 $\pm$ .16
 & .59 $\pm$ .17 & .75 $\pm$ .17 \\

Qwen-3-32B
 & .59 $\pm$ .14 & .82 $\pm$ .09
 & .72 $\pm$ .11 & .87 $\pm$ .07
 & .58 $\pm$ .07 & .65 $\pm$ .03
 & .66 $\pm$ .12 & .68 $\pm$ .11
 & .65 $\pm$ .13 & .80 $\pm$ .12 \\
 
GPT-4o
 & .46 $\pm$ .14 & .74 $\pm$ .13
 & .60 $\pm$ .13 & .82 $\pm$ .08
 & .38 $\pm$ .17 & .38 $\pm$ .14
 & .48 $\pm$ .08 & .52 $\pm$ .06
 & .51 $\pm$ .15 & .69 $\pm$ .19 \\
 
GPT-OSS-120B
 & .64 $\pm$ .11 & .82 $\pm$ .06
 & .75 $\pm$ .16 & .88 $\pm$ .08
 & .65 $\pm$ .22 & .67 $\pm$ .07
 & .74 $\pm$ .17 & .76 $\pm$ .12
 & .70 $\pm$ .17 & .81 $\pm$ .11 \\
 
GPT-5.2
 & .74 $\pm$ .18 & .88 $\pm$ .07
 & .75 $\pm$ .17 & .87 $\pm$ .10
 & .83 $\pm$ .14 & .80 $\pm$ .11
 & .84 $\pm$ .15 & .80 $\pm$ .15
 & .77 $\pm$ .16 & .85 $\pm$ .11 \\

GPT-5.4
 & \textbf{.83 $\pm$ .14} & \textbf{.90 $\pm$ .06}
 & \textbf{.82 $\pm$ .15} & \textbf{.90 $\pm$ .06}
 & \textbf{.81 $\pm$ .09} & \textbf{.79 $\pm$ .08}
 & \textbf{.80 $\pm$ .14} & \textbf{.84 $\pm$ .13}
 & \textbf{.82 $\pm$ .14} & \textbf{.87 $\pm$ .09} \\
 
\bottomrule
\end{tabular}
}
\end{table}

\subsection{Evaluation Metrics}
We evaluate the performance of the systems using Normalized Discounted Cumulative Gain (NDCG) at cutoff $k \in \{10,100\}$ \cite{wang2013theoretical}. Given a ranked list of sentences, the Discounted Cumulative Gain (DCG) at rank $k$ is defined as
\begin{equation*}
\mathrm{DCG}@k = \sum_{i=1}^{k} \frac{2^{\mathrm{rel}_i} - 1}{\log_2(i + 1)},
\end{equation*}
where $\mathrm{rel}_i$ denotes the ground-truth relevance label of the sentence ranked at position $i$. The Ideal DCG is computed by sorting the candidates in decreasing order of relevance.
% \begin{equation}
% \mathrm{IDCG}@k = \sum_{i=1}^{k} \frac{2^{\mathrm{rel}_i^{\ast}} - 1}{\log_2(i + 1)},
% \end{equation}
% where $\mathrm{rel}_i^{\ast}$ is the relevance label at position $i$ in the ideal ranking.
Finally, NDCG@k is defined as DCG@k over IDCG@k.
% \begin{equation}
% \mathrm{NDCG}@k = \frac{\mathrm{DCG}@k}{\mathrm{IDCG}@k}.
% \end{equation}

\section{Results}
The results of our experiments are reported in Table \ref{tab:full_results}. The top portion of the table shows results of prior work \cite{savelka2021discovering,libal2025manual}. The middle section presents results for newly evaluated ModernBERT variants. The ModernBERT models consistently match the results of the prior work performed with earlier BERT-based models across most configurations. Further, we observe that variants that incorporate broader contextual information, such as sp2ctx and sp2snt+ctx, surprisingly degrade performance, especially in large sparse queries.

LLM-based approaches demonstrate strong overall performance, particularly for dense queries. Zero-shot with GPT-4o and LLaMA models yield results comparable to encoder-based methods. Among all evaluated configurations, GPT-5.4 with probabilistic scoring achieves the highest overall NDCG@10 and NDCG@100 scores, beating reported state-of-the-art~\cite{libal2025manual}.

\section{Discussion}
\paragraph{Context Expansion} 
Across both encoder-based rankers and LLM-based prompting, increasing the available context does not yield reliable gains in ranking quality and can substantially degrade performance. For encoder models, \textit{ModernBERT sp2snt} achieves NDCG@10 of $0.53$ in the large sparse subset, but drops to $0.51$ with \textit{sp2ctx} and collapses to $0.28$ with \textit{sp2snt+ctx} (NDCG@100: $0.51 \rightarrow 0.49 \rightarrow 0.35$). 

Expanded inputs reduce the effective signal-to-noise ratio. Relevance is often driven by a small number of legally meaningful localized signals \cite{lin2020pretrained}, and longer inputs simply enlarge the search space without providing additional supervision to suppress distracting content. This burden is particularly acute in the large sparse subset, where high- and certain-value sentences are rare. The effect is especially pronounced when the additional text is accompanied by structural markers only for fine-tuning (e.g., \verb|<query>qry</query>|) intended to highlight the term span in the statutory provision (i.e., \textit{sp+qry2snt} mode).

Unlike established control tokens such as \texttt{[CLS]} or \texttt{[SEP]} which are present in pre-training and therefore acquire stable functional semantics, these fine-tuning-only tokens are not reliably grounded in the model's pre-trained inductive biases. Consistent with this, \textit{sp+qry2snt} underperforms \textit{sp2snt} on the large sparse subset (NDCG@10: $0.53 \rightarrow 0.34$; NDCG@100: $0.51 \rightarrow 0.32$).

\paragraph{Advantages of Decoder-Only Models}
Decoder-only models consistently outperform encoder-based rankers in our experiments, with the largest gains on the most challenging large sparse queries. Among all evaluated models, \textbf{GPT-5.4} achieves the strongest overall performance ($0.82/0.87$ NDCG@10/100), with consistent gains across all subsets including the difficult LgSp setting ($0.81/0.79$). GPT-5.2 remains highly competitive ($0.77/0.85$), and the open-weight GPT-OSS-120B also delivers strong results ($0.70/0.81$), outperforming smaller open-weight models such as Qwen-3-32B ($0.65/0.80$), LLaMA-4-Scout-17B ($0.59/0.75$), and LLaMA-3.3-70B ($0.58/0.75$). GPT-4o lags behind the other prompted models~($0.51/0.69$), suggesting that model scale and instruction-following quality are important factors in this task. 

We hypothesize that the gains of larger models stem from stronger semantic priors of instruction-tuned LLMs and their ability to jointly reason over statutory language, candidate sentences, and surrounding legal context, without task-specific fine-tuning or heavily engineered input representations. Specifically, these models demonstrate a nuanced capacity to look beyond superficial lexical overlap, successfully distinguishing procedurally dense boilerplate from sentences that offer genuine interpretive utility.

The advantage of prompting-based approaches is most evident in time-to-results. GPT-5.4 processes the full dataset in approximately 5 minutes end-to-end, whereas \textit{ModernBERT sp2snt} incurs a substantially higher upfront training cost. On Google Colab with $3\times$ A100 GPUs, six-fold training for 10 epochs~(batch size 8) requires roughly 3 hours in total; once trained, however, \textit{ModernBERT sp2snt} can score the full dataset in only 2--3 minutes. Open-weight models via Groq offer competitive performance at lower cost, though they require more aggressive output repair due to higher rates of JSON formatting errors.

\section{Conclusion}
Our results yield three notable takeaways. First, ModernBERT largely matches prior BERT-family performance in comparable input settings, indicating that architectural improvements alone do not guarantee gains when signals and the target task remain unchanged. Second, adding broader context does not yield reliable improvements. Both encoder-only- and decoder-only-based approaches degrade substantially when inputs are expanded to include paragraph-level context or long document context, particularly for large sparse queries where relevant evidence is rare. Third, decoder-only prompting, especially probability-based scoring, achieves the strongest overall performance (consistent with \cite{libal2025manual}).

Our findings suggest that effective explanatory sentence ranking depends less on providing more text and more on reliably amplifying task-relevant cues while suppressing noise, and that modern instruction-tuned LLMs offer a strong effectiveness--engineering trade-off for this problem.

For future work, recent hybrid systems demonstrate strong effectiveness by combining sparse and dense candidate generation with a learned re-ranker (e.g., BM25 and dense FAISS retrieval) to shortlist candidates, followed by an LLM-based cross-encoder for final ordering \cite{sager2025deep}. Inspired by this design, a practical extension of our work is a two-stage pipeline that (i) uses fast sparse encoder-only retrieval to produce a small candidate set, and (ii) applies a decoder-only re-ranker with probabilistic (expected-value) scoring to break ties and refine the top ranks. Additionally, our dataset covers U.S. Code and a specific case-law sources. Extending evaluation to additional jurisdictions, statutory sources, and domains (including non-English settings) would shed light on how well our findings transfer.

%% The declaration on generative AI comes in effect
%% https://ceur-ws.org/GenAI/Policy.html
\section*{Declaration on Generative AI}

During the preparation of this work, the authors used ChatGPT to: grammar and spelling check, paraphrase, and reword. After using this tool, the authors reviewed and edited the content as needed and assume full responsibility for the content of the publication.

%%
%% Define the bibliography file to be used
\bibliography{main}

\appendix

\end{document}